\documentclass[pre,twocolumn,10pt,aps,eqsecnum,showpacs,natbib]{revtex4-1}

\usepackage[dvips]{graphicx}
\usepackage{textcomp}
\usepackage{amssymb}

\usepackage{times} 
\usepackage{graphicx,color}

\mathchardef\bigtilde="0365

\begin{document}

\title{Large-order aspects of the $\delta$ expansion in low-dimensional Ising models}
\author{Hirofumi Yamada}\email{yamada.hirofumi@it-chiba.ac.jp}
\affiliation{%
Division of Mathematics and Science, Chiba Institute of Technology, 
\\Shibazono 2-1-1, Narashino, Chiba 275-0023, Japan}

\date{\today}

\begin{abstract}
{We investigate the large order aspects of the $\delta$-expansion under the estimation procession of the critical quantities.   As illustrative examples, we revisit the one-dimensional Ising model for the analytic study and the two-dimensional square Ising model in the high-temperature phase for the numerical experiment to large orders.  In both models, the proposed fundamental base on which the estimation protocol should be constructed is investigated in details and confirmed to be valid.  In the square lattice model, we present a new protocol for the estimation of critical exponents and temperature.}
\end{abstract}

\pacs{11.15.Me, 11.15.Pg, 11.15.Tk}

\maketitle
\section{Introduction}
Over an year ago, we studied an Ising model on a cubic lattice \cite{yam} by use of the $\delta$ expansion \cite{yam2}.   The study is based upon the high temperature expansion and the resulting series rewritten in the inverse of a mass squared $M$ is transformed by the $\delta$-expansion which played a crucial role in the estimation of critical quantities.  However the foundamentals of the estimation protocol, which will be explicitly stated in the next section, have not  been thoroughly investigated so far.   
This paper deals with the issue in enough depth via the study of Ising models at one- and two-dimensions where the models are solved and high temperature expansion is available to very large orders.   
Another purpose of this paper is to improve the estimation protocol of critical quantities employed in Ref. \cite{yam}.  We will give a renewed procedure and confirm it works better in the low dimensional models, where the estimation of $\nu$, $\gamma/2\nu$ and $\beta_{c}$ will be presented on the square lattice.  The new protocol would be applicable to high dimensional models and even field theoretic models.  Before the application to these more interesting models, it needs a detailed test on well understood models.

The Ising model we consider is a pure one without applied forces, specified by the action\begin{equation}
\beta H=-\beta\sum_{<i,j>}s_{i}s_{j},\quad s_{i}^2=1
\end{equation}
where $\beta$ denotes the inverse temperature and $<i,j>$ means the pair of nearest neighbor sites.  It is well known that a one-dimensional model is in the one phase and the square lattice model undergoes a second order phase transition at $\beta_{c}=\log(1+\sqrt{2})/2$.  
One often discusses critical behaviors of thermodynamic quantities in terms of 
$
\tau=1-\beta/\beta_{c}
$ \cite{camp}
where $\beta_{c}$ stands for the critical temperature.  In the high temperature region, in contrast, thermodynamic quantities are expanded in $\beta$.   As the temperature approaches from a hot region to the vicinity of the critical point, the effective parameter changes from $\beta$ to $\tau$.  We feel that this exchange in the basic descriptive parameter is somewhat strange.    There exists, however, a sole parameter effective in the high-temperature phase.  We note that the second order transition is activated by the divergence of the correlation length.  The physics is controlled by correlation length $\xi$.   The critical point is specified as the point where $\xi$ is divergent and this means that, irrespective of specific value of $\beta_{c}$ which is of non-universal, the critical point is universally specified by the point of $\xi=\infty$ or the massless point $M=0$ where $M$ should be understood as suitably related to $\xi$.   If one describes the system in $M$,  even the temperature is a function of $M$ (one can obtain $\beta(M)$ by the inversion of $M(\beta)$) and the critical temperature is given by the limit
\begin{equation}
\beta_{c}=\lim_{M\to 0}\beta(M).
\end{equation}
The high-temperature expansion corresponds to the $1/M$ expansion.  From the large mass side, we access the critical point with the crucial help of the $\delta$-expansion \cite{yam2}.  By making use of the fact that the critical point is at the massless point, we will explore a new way of handling high-temperature expansion.  

The paper is organized as follows:  In Sec. 2, the $\delta$-expansion is reviewed.  We posit a base on which the estimation protocol is constructed.  In Sec. 3, a one-dimensional case is taken up to analytically study the $\delta$-expansion to arbitrary orders.   In Sec. 4, a two-dimensional model on the square lattice is numerically revisited.  The parametric extension proposed in Ref. \cite{yam} is re-formulated by the linear differential equation (LDE) with constant coefficients.  The spectrum of exponents in the series expansion in the vicinity of the critical point is given by the spectrum of roots of the characteristic equation of the LDE satisfied by the thermodynamic quantities.  The critical temperature and the amplitudes appear as the integration constants.    Studies to quite large orders tell us that the computational base under the $\delta$-expansion is numerically confirmed.  Then, under a new protocol to be proposed, we demonstrate that the estimation becomes more accurate.  In the final section, we give concluding remarks.

\section{Brief review of the $\delta$-expansion}
The $\delta$-expansion originates in the dilatation of the scaling region.   With the dilatation rate $(1-\delta)^{-1}$ $(0\le \delta\le 1)$, the change of the parameter from $M$ to $1/t$ by $M=(1-\delta)/t$ in a given thermodynamic quantity $\Omega(M)$ induces a dilatation transform as follows:   Consider the large mass expansion truncated at order $N$, $\Omega_{N}(M)=\sum_{n=0}^{N}a_{n}/M^n$.  Then, truncate the expansion in $\delta$ in $1/M^{n}=1/((1-\delta)/t)^{n}$ such that the sum of the expansion order of $\delta$ and the order of $M^{-1}$ is less than or equal to $N$ but not over $N$.   This specific truncation rule yields
\begin{eqnarray}
M^{-n}&=& ((1-\delta)/t)^{-n}\sim t^{n}(1+n\delta+\cdots\nonumber\\
& &+\frac{N!}{(n-1)!(N-n+1)!}\delta^{N-n}).
\end{eqnarray}
Then, taking the limit $\delta\to 1$ which means formally the infinite rate dilatation, we obtain
\begin{equation}
M^{-n}\to C_{N,n}t^n
\label{delta1}
\end{equation}
where $C_{N,n}$ denotes the binomial coefficient
\begin{equation}
C_{N,n}=\frac{N!}{n!(N-n)!}.
\end{equation}
Thus,  
the results of the $\delta$ expansion denoted by $D_{N}[\Omega_{N}]$ or $\bar\Omega_{N}$ are obtained as a function of $t$,
\begin{equation}
D_{N}[\Omega_{N}]=\bar\Omega_{N}(t)=\sum_{n=0}^{N} C_{N,n}a_{n}t^n.
\end{equation}
On the other hand, for a general power of $M$ we use the analytic extension of the $\Gamma$ function giving
\begin{equation}
D_{N}[M^{\alpha}]=C_{N,-\alpha}t^{-\alpha}=\frac{\Gamma(N+1)}{\Gamma(-\alpha+1)\Gamma(N+\alpha+1)}t^{-\alpha}.
\label{delta2}
\end{equation}
One can deduce two useful results:  By setting $\alpha=1,2,3\cdots$, we have 
\begin{equation}
D_{N}[M^{\alpha}]=0,\quad \alpha=1,2,3,\cdots.
\label{regular}
\end{equation}
By the expansion of (\ref{delta2}) in $\alpha$, it follows, for example at order $\alpha^{1}$,
\begin{equation}
D_{N}[\log M]=-\log t-\sum_{n=1}^{N}\frac{1}{n}.
\label{deltalog}
\end{equation}
By comparison of both sides of (\ref{delta2}) at order $\alpha^{k}$, one has $D_{N}[(\log M)^k]$.  Further, 
if one sets $\alpha=n+\epsilon$ with infinitesimal $\epsilon$, expansion in $\epsilon$ provides a similar formula for $M^n$ times logarithms.

Assuming the power-like behavior of thermodynamic function $\Omega(M)$ in the scaling region, let us explain the effect of $\delta$-expansion.  Series here is $1/M$ expansion truncated at order $N$.  
Then, our task is to obtain the critical information of the supposed behavior
\begin{equation}
\Omega(M)\sim \Omega_{0}+c_{1}M^{\lambda_{1}}+c_{2}M^{\lambda_{2}}+\cdots.
\label{scaling1}
\end{equation}
Here, $0<\lambda_{1}<\lambda_{2}<\cdots$ and $\lim_{M\to 0}\Omega(M)=\Omega_{0}$.  There are other types of critical behaviors such that $\Omega(M)\sim c_{1}M^{-\lambda_{1}}+c_{2}M^{-\lambda_{2}}+\cdots$  $(\lambda_{1}>\lambda_{2}>\cdots)$ and $\Omega(M)\sim c_{0}\log M+\cdots$.  These are typical in the study of phase transition as these two  represent the divergences of thermodynamic functions. 

Let us explain one of the advantages in the $\delta$-expansion, taking the scaling form (\ref{scaling1}) as an example.  The construction via $\delta$-expansion gives
\begin{equation}
D_{N}[\Omega]=\Omega_{0}+c_{1}C_{N,-\lambda_{1}}t^{-\lambda_{1}}+c_{2}C_{N,-\lambda_{2}}t^{-\lambda_{2}}+\cdots.
\label{delta_scale}
\end{equation}
Note that order $N$ of large mass expansion is included in the coefficients in (\ref{delta_scale}).  
The first term is most important and the rest acts as the corrections to the first one.  In the limit $N\to\infty$, respective terms of the corrections tend to vanish since
\begin{equation}
C_{N,-\lambda_{n}}\to \frac{N^{-\lambda_{n}}}{\Gamma(1-\lambda_{n})}.
\label{scale_c1}
\end{equation}
This represents the scaling of the amplitude $c_{1}C_{N,-\lambda_{n}}$ at large enough order.  Thus, as $N\to \infty$,   
\begin{equation}
C_{N,-\lambda_{n}}t^{-\lambda_{n}}\to \frac{(N t)^{-\lambda_{n}}}{\Gamma(1-\lambda_{n})},
\label{scale_c}
\end{equation}
and the $n$th-order correction tends to zero for fixed $t$.  
This is quite convenient for the estimation of $\Omega_{0}$.   

The above result (\ref{scale_c}) suggests us that for $\Omega_{N}(M)$ where $N$ denotes the order of large $M$ expansion,
\begin{equation}
\lim_{N\to \infty}D_{N}[\Omega_{N}](t)=\Omega_{0},
\label{delta_limit}
\end{equation}
{\it in some region $I$ of $t$}.  
As mentioned in the previous section, this manifests a major characteristic property of $\delta$ expansion and plays a foundamental role in estimating the critical quantities.   Though formal reasoning supports (\ref{delta_limit}),
 if we inspect the terms in the large mass expansion, we find $C_{N,n}t^{n}\to (Nt)^n/n!$ which diverges to infinity as $N\to \infty$ even when $t$ is small.  Though previous work \cite{yam} provided confirmed results on the point, the study is limited to $25$th order and it is as yet non-trivial whether the fundamental result (\ref{delta_limit}) remains valid to large enough orders.  We shall deeply investigate the point in two solvable models in the following sections.

\section{One dimensional Ising model}
In the one-dimensional Ising model, it is well known that the exact correlation length $\xi$ is given by $\xi^{-1}=-\log(\tanh \beta)$.  
Then, the mass squared $M$ defined by the zero momentum limit of correlation function is obtained by $M=2(\cosh \xi^{-1}-1)$.  We note that $M$ and $\xi$ do not obey the relation $M= \xi^{-2}$ but satisfy $M=\xi^{-2}+\xi^{-4}/12+\cdots$ ($\xi\gg 1$) \cite{comment} and inversion gives
\begin{equation}
\beta=\frac{1}{4}\log(1+4x),
\end{equation}
where
\begin{equation}
x=\frac{1}{M}.
\end{equation}
The function $\beta(x)$ has a branch point at $x=-1/4$, and the large mass expansion,
\begin{equation}
\beta=\frac{1}{4}\sum_{n=1}^{\infty}\frac{(-1)^{n-1}}{n}(4x)^n=:\beta_{>},
\label{largem1}
\end{equation}
is valid within the convergence region $-1/4<x<1/4$.  
Here we have introduced the notation $\beta_{>}$, which stands for the inverse temperature at small $x$.  
At large $x$, we make a division such as $\log(1+4x)=\log(4x)+\log(1+1/4x)$ and obtain the expansion,
\begin{equation}
\beta=\frac{1}{4}\log(4x)+R=:\beta_{<},
\label{critical1}
\end{equation}
where the notation $\beta_{<}$ means the expansion of $\beta$ near the critical point and $R$ denotes the regular part given by
\begin{equation}
R=\frac{1}{4}\sum_{n=1}^{\infty}\frac{(-1)^{n-1}}{n}(4x)^{-n}.
\end{equation}
Since $\beta$ shows logarithmic divergence in the $M\to 0$ limit, $\beta_{c}=\infty$.  We call the behavior (\ref{critical1}) a critical behavior.  From (\ref{deltalog}), we find
\begin{equation}
D_{N}[\beta_{<}]=\frac{1}{4}\Big(\log(4t)+\sum_{k=1}^{N}\frac{1}{k}\Big)+D_{N}[R]=:\bar\beta_{N<}.
\end{equation}

Now, we consider $D_{N}[\beta_{N>}]=\bar\beta_{N>}$ where $\beta_{N>}$ stands for the truncation of $\beta_{>}$ to $N$th order such that $\beta_{N>}=\frac{1}{4}\sum_{n=1}^{N}\frac{(-1)^{n-1}}{n}(4x)^n$.  We then have
\begin{eqnarray}
\bar\beta_{N>}&=&\frac{1}{4}\sum_{n=1}^{N}\frac{(-1)^{n-1}}{n}C_{N,n}(4t)^n\nonumber\\
&=&\frac{1}{4}\int ^{1}_{1-4t}\frac{1-z^N}{1-z}dz.
\label{beta_closed}
\end{eqnarray}
The above integral representation shows that the region where $\bar\beta_{N>}$ remains valid in the $N\to \infty$ limit is $|1-4t|<1$ ($0<t<1/2$).   Note that the center of the region $I=(0,1/2)$ is $t=1/4$.  This indicates that the correct expansion should be inside region $I$ and not at $t=0$ on the boundary of $I$.   From (\ref{beta_closed}), we obtain the expansion around the center $t=1/4$:
\begin{equation}
\bar\beta_{N>}=\frac{1}{4}\Big(\log(4t)+\sum_{k=1}^{N}\frac{1}{k}+\sum_{n=N+1}^{\infty}\frac{(1-4t)^n}{n}\Big).
\label{beta_expansion}
\end{equation}
From the comparison of (\ref{critical1}) and (\ref{beta_expansion}), we find
\begin{equation}
D_{N}[R]=\sum_{n=N+1}^{\infty}\frac{(1-4t)^n}{n}.
\end{equation}
Thus we cannot use $D_{N}[x^{-n}]=D_{N}[M^n]=0\,(n=1,2,3,\cdots)$ inside the infinite sum, though the terms of the form $t^{-n}$ are absent in the right-hand-side (The expansion of the right-hand-side around $t=\infty$ is impossible).  The result of $\delta$-expansion reads $D_{N}[R]=O(\epsilon^{N+1})$ where $\epsilon=t-1/4$ and the correction to the logarithmic scaling shows rapid suppression at large $N$ in the neighborhood of $t\sim 1/4$.  In other words, as $N$ tends to infinity,  $D_{N}[R]$ becomes negligible and 
$\bar\beta\to \frac{1}{4}(\log(4Nt) +\gamma_{E})\quad (0<t<1/2)$.  
Thus, we are convinced that the divergence of the coefficient in the $N\to \infty$ reflects that the point $t=0$ is just on the boundary of the convergence region $I=(0,1/2)$.

The $N\to \infty$ limit of $\bar\beta_{N>}$ diverges for any $t\in I$, and this is due to the presence of the logarithmic singularity in the original function as $M\to 0$.  This case shows a variant of (\ref{delta_limit}) in the case of a logarithmically divergent limit.  We are able to capture the behavior of the logarithmic divergence in $\bar\beta_{N>}$ via numerical analysis.   See Fig. 1 where $\bar\beta_{N>}$ is depicted at $N=10$.
\begin{figure}[h]
\centering
\includegraphics[scale=0.95]{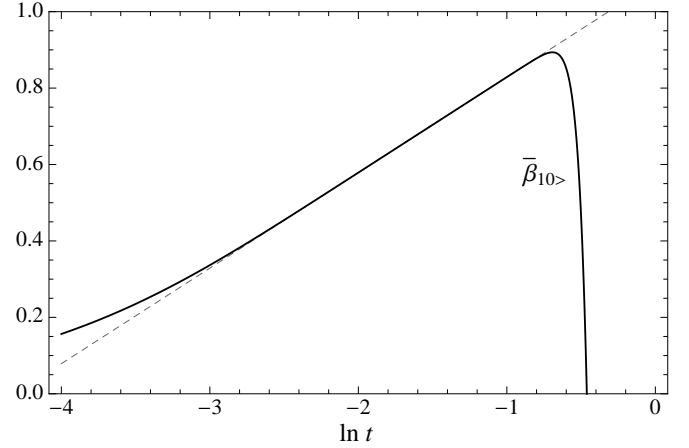}
\caption{Plot of $D_{N}[\beta_{10>}]=\bar\beta_{10>}$ against $\log t=\ln t$.  The broken gray line indicates the exact asymptotic behavior $(\log(4t)+\sum_{k=1}^{10}1/k)/4$.}
\end{figure}
By study of the derivative $\beta^{(1)}$, we find further evidence of logarithmic behavior and the coefficient of the log, $1/4$:  We find directly from (\ref{beta_closed}) by the differentiation with respect to $d/d\log t$,
\begin{equation}
\bar\beta_{N>}^{(1)}=\frac{d\bar\beta_{N>}}{d\log t}=\frac{1}{4}(1-(1-4t)^N),
\label{3_10}
\end{equation}
and the exact validity of (\ref{delta_limit}) is stated as
\begin{equation}
\lim_{N\to \infty}\bar\beta_{N>}^{(1)}=\frac{1}{4},\quad t\in I.
\end{equation}
For convenience, we show the plots of $\bar\beta_{N>}^{(1)}(t)$ at $N=10,45,$ and $100$ in Fig. 2.  
\begin{figure}
\centering
\includegraphics[scale=0.9]{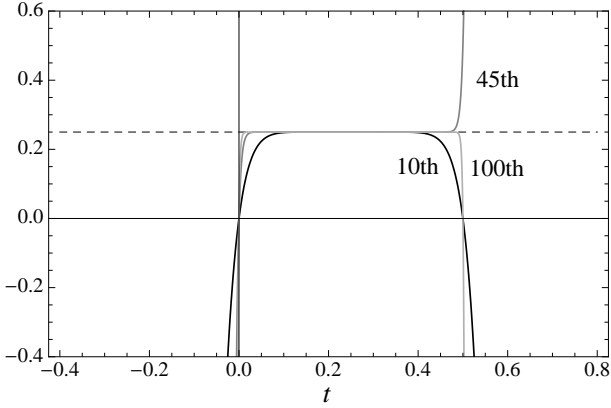}
\caption{Plots of $\bar\beta_{N}^{(1)}$ at $N=10,45,100$.  The dashed gray line indicates the limit $1/4$. At every orders, the unique stationary point gives the exact limit.}
\end{figure}
In accord with (\ref{3_10}), the functions imply the step-like function, which is almost flat over the region $I=(0,1/2)$.   At finite $N$, the unique stationary point is always realized at $t=1/4$ and gives an exact value of $\lim_{N\to \infty}\bar\beta_{N>}^{(1)}$, which agrees with the coefficient of $\log t$.  It is also instructive to note that the higher order derivatives satisfy $\bar\beta_{N>}^{(\ell)}|_{t=1/4}=0$ $(\ell=2,3,4,\cdots, N)$.

\section{Two dimensional Ising model}
\subsection{High and low temperature expansions}
In a one-dimensional Ising model, the scaling behavior is simple and allows exact analysis.  In the two-dimensional model, a complexity related to the critical behaviors appears.   Most of our investigation is based on the exact solution, and the model serves as a good testing ground for understanding large order aspects of $\delta$ expansion.

The exact result on which our study based is a relation between the correlation length and inverse temperature \cite{mont},
\begin{equation}
\xi^{-1}=-\log\tanh\beta-2\beta,
\label{2dxi}
\end{equation}
where $\xi$ stands for the correlation length extracted in the large separation limit of a two-point function along one of the two axes.  
From above, we obtain a well-known result:
\begin{equation}
\beta_{c}=\frac{1}{2}\log(1+\sqrt{2})=0.44068679\cdots.
\end{equation}
In addition we have
\begin{equation}
\xi^{-1}=4\beta_{c}\tau+2\sqrt{2}\beta_{c}^2\tau^2+4\beta_{c}^3\tau^3+\cdots
\end{equation}
where $\tau=1-\beta/\beta_{c}$, and we find
\begin{equation}
\nu=1.
\end{equation}

The mass squared $M$ corresponding to $\xi$ is given by the transformation $M=2(\cosh \xi^{-1}-1)$ \cite{tarko}.  From the straightforward expansion
\begin{equation}
M=16\beta_{c}^2\tau^2+16\sqrt{2}\beta_{c}^3\tau^3+\cdots,
\end{equation}
we obtain
\begin{equation}
\beta_{c}\tau=\frac{1}{4}M^{1/2\nu}+\cdots,\quad (\nu=1),
\label{tau_scale}
\end{equation}
and arrive at
\begin{eqnarray}
\beta_{<}&=&\beta_{c}-\frac{1}{4}M^{1/2\nu}\Big(1-\frac{1}{24}M+\cdots\Big)+R,\label{crit_2d_a}\\
R&=&\frac{1}{16\sqrt{2}}M-\frac{3}{512\sqrt{2}}M^{2}+\cdots,
\label{crit_2d}
\end{eqnarray}
where $R$ denotes the regular part.  As well-known, the two-dimensional model has no confluent singularity, and the structure of the expansion is rather simple.   In our analysis, we do not intend to use the above simplicity.  We just assume the power-like expansion
\begin{eqnarray}
\beta_{<}&=&\beta_{c}-A_{1}M^{p_{1}}-A_{2}M^{p_{2}}-\cdots\label{beta_scale1}\\
&=&\beta_{c}-A_{1}M^{p_{1}}\Big(1+A_{2}/A_{1}M^{p_{2}-p_{1}}+\cdots\Big)
\label{beta_scale2}\nonumber
\end{eqnarray}
where $0<p_{1}<p_{2}<\cdots$.  

The large mass expansion of $\beta$ is obtained to an arbitrary large order from (\ref{2dxi}).  To the first several orders it reads
\begin{equation}
\beta_{>}=x-4x^2+\frac{58}{3}x^3-104x^4+\frac{3006}{5}x^5-\cdots.
\label{4_10}
\end{equation}
We briefly comment on the analytic extension of original $\beta_{N>}(x)$ on the complex $x$-plane using Pad\'e approximates:  The poles and zeros line up side by side on the negative real axis.  The pole of the largest absolute value extends toward $-\infty$, and the pole of the smallest absolute value seems to be converging to $-1/8$.  From conventional knowledge, this kind of set of poles and zeros indicates the cut.  The implied convergence radius of $1/M$ expansion is $1/8$ and this agrees with the simple study of the coefficient ratio.

\subsection{Effective region of $\delta$-expansion}
In the one-dimensional model, the convergence region of original $1/M$ expansion is $|x|<1/4$ and the $\delta$-expanded counterpart reads $|t-1/4|<1/4$.  It is as if the region of convergence is shifted right by $1/4$.  If $t$ is negative and small in the absolute value, $\bar\beta_{N}$ diverges in the $N\to \infty$ limit.  This may sound strange, since the series in powers of $x$ is effective around $x=0$ and then the expansion in $t$ would ordinarily include the origin within the convergence circle.  

Before proceeding to estimate critical quantities in the two-dimensional model, we would like to elucidate the point in the square lattice.   From (\ref{4_10}), we obtain
\begin{equation}
\bar\beta_{N>}=C_{N,1}t-4C_{N,2}t^2+\frac{58}{3}C_{N,3}t^3-104C_{N,4}t^4+\cdots,
\end{equation}
\begin{figure}
\centering
\includegraphics[scale=0.9]{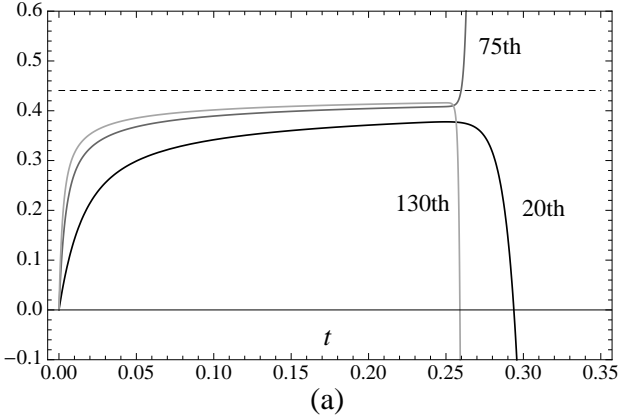}
\includegraphics[scale=0.9]{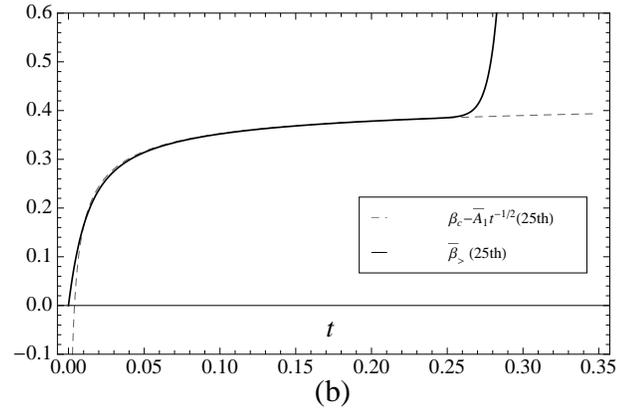}
\caption{(a)\, The plots of $\bar\beta_{N>}$ at orders $20$th, $75$th and $130$th.  Dashed line indicates $\beta_{c}$.  (b)\,The plots of $\bar\beta_{N>}$ (solid line) and $\beta_{c}-\bar A_{1}t^{-1/2}$ (dashed gray line) at $N=25$.  Here we note that $\bar A_{1}=A_{1}C_{N,-1/2}=(1/4)C_{N,-1/2}$.}
\end{figure}
We first show in Fig. 3 the plots of $\bar\beta_{N>}$ at various orders.  
As seen in the upper graph (a), $\bar\beta_{N>}$ shows slow convergence to $\beta_{c}$.  This is due to the presence of the corrections, especially the leading one $\sim t^{-1/2}$.  Actually, the simultaneous plot of $\bar\beta_{N>}$ and $\beta_{c}-\bar A_{1}t^{-1/2}$ ($A_{1}=1/4$ and (\ref{delta2}) are used) at $N=25$ shown in Fig. 3 (b) reveals that the two graphs are in excellent agreement.  Thus, we find that $\bar\beta_{N>}$ precisely approximates $\bar\beta_{N<}$ in its simplest form and, as long as $C_{N, -1/2}\to 0$ in the $N\to \infty$ limit, $\bar\beta_{N>}\to \beta_{c}$ in the region $I=(0,1/4)$.   

Figure 4 shows the plots of $(1+2d/d\log t)\bar\beta_{N>}$ at various orders.  The coefficient $2$ multiplied to $d/d\log t$ is chosen from the asymptotic form of $\beta$ near the critical point $\beta_{<}(x)\sim \beta_{c}+const\times x^{1/2}$ (see (\ref{crit_2d_a})), i. e., the inverse of the critical exponent $1/2$ is used.  The value makes the leading correction, the term $const\times x^{1/2}$, be canceled between $\beta$ and $\beta^{(1)}=d\beta/d\log x$.   Thus the correction is highly suppressed in the combination.  
As a result, $(1+2d/d\log t)\bar\beta_{N>}$ shows a clear convergence trend over the whole region of $I=(0,1/4)$. 
\begin{figure}[t]
\centering
\includegraphics[scale=0.9]{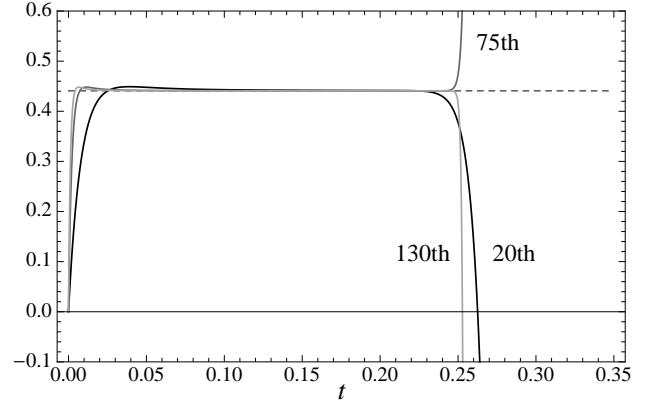}
\caption{Plots of $\bar\beta_{N>}+2\bar\beta_{N>}^{(1)}$ at $N=20$, $75$ and $130$.  Dashed line indicates $\beta_{c}$.}
\end{figure}

We can say that the region supporting (\ref{delta_limit}) thus indeed exists for the functions of interest.  Explicitly, the region is confirmed to be $I=(0, 1/4)$ approximately.  The point is, as in a one-dimensional model, that the origin $t=0$ seems to be excluded from this region.   To confirm it in a more conclusive, we compute $\exp(\bar\beta_{N>})$ at large orders $N=300$ and $301$ over the region $(-0.3,0.3)$.  The result is plotted in Fig. 5.  The effective region where $\bar\beta_{N>}$ takes a finite value is limited between $t\sim 0$ and $\sim 0.25$.   In other regions, we find $\bar\beta_{N>}\to \pm \infty$.
\begin{figure}
\centering
\includegraphics[scale=0.9]{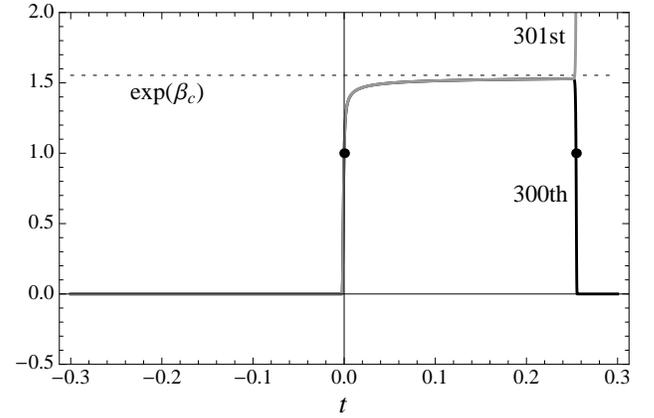}
\caption{Plots of $\exp(\bar\beta_{N>})$ (dotted) at $N=300$ (black) and $301$ (gray) as functions of $t$.  The filled blobs signal points $(0, 1)$ and $(0.254378, 1)$ at which $\bar\beta_{N>}=0$.}
\end{figure}

The above feature is investigated from another angle by using Pad\'e approximants.  Since $\bar\beta_{N>}$ and the derivatives converge respectively to $\beta_{c}$ and zero, the diagonal approximants denoted as $[n/n]$ ($N=2n$) are the most appropriate choice in the  Pad\'e table.  Figures. 6 (a)-(b) show the poles and zeros of the $[n/n]$ element of $\bar\beta_{N>}$ and $\bar\beta_{N>}^{(1)}$ at $N=24$, $50$ and $300$.  From Fig. 6, we find, in spite of coefficients of $\bar\beta_{N}(t)$ and $\bar\beta^{(1)}_{N}(t)$ being dependent on the order, that both shapes of distributed poles and zeros seem to converge to circular ones, though the one for poles is chipped.  The shape of the pole set is apparently recognized as the circle even at rather low orders and the convergence is not slow.  In contrast, the shape of the zero set changes very slowly for $\bar\beta_{N>}$ though the reason behind it is not clear to us.   

At $N=300$ for $\bar\beta_{N>}$, the points at the left end of pole sets have the coordinate $-0.1265831 \pm 
 0.0015247\, i$, and the points at the right chipped end have $0.0551246\pm  
 0.1133955\,i$.  As for $\bar\beta_{N>}^{(1)}$ at the same order, the points at the left end of pole sets have the coordinate $-0.1259443\pm  
 0.0015187\, i$ and the points at the right end have $0.0551476\pm 
 0.1129012\,i$.  The respective norms are $0.1265923$ and $0.1260843$ for $\bar\beta_{N>}$ and $0.1259534$ and $0.1256501$ for $\bar\beta_{N>}^{(1)}$.  The deflection angles for the upper right end points are, respectively, $\pi/3\cdot 1.0679062$ and $\pi/3\cdot 1.0661059$ for $\bar\beta_{N>}$ and $\bar\beta_{N>}^{(1)}$.   From these results and those at lower orders, it is strongly suggested that in the infinite order the both sets of poles for $\bar\beta_{N>}$ and $\bar\beta_{N>}^{(1)}$ are dense and form the chipped circle with the radius $1/8$, the center being at the origin and upper end points having deflection angle $\pi/3$.  Both zero-point sets in the $N\to \infty$ limit form dense sets of the circle with radius $1/8$ and center at $t=1/8$.  Then, if a $t$-series obtained by the $\delta$-expansion would be a series expanded around $t=0$, it must be effective "inside" of the chipped pole-circle and not effective for $Re[t]>1/8$.  But it is not actually the case as manifest in Fig. 5.   Thus, we conclude that effective region of $\delta$-expansion is shown by the zero-circle, not by the pole-circle.  The origin $t=0$ is actually at the boundary of the effective region and this is in accord with the fact that the coefficient of $t^k$ diverges as $N\to \infty$; that is, the $\delta$-expansion does not provide expansion around $t=0$. 
\begin{figure}[t]
\centering
\includegraphics[scale=0.85]{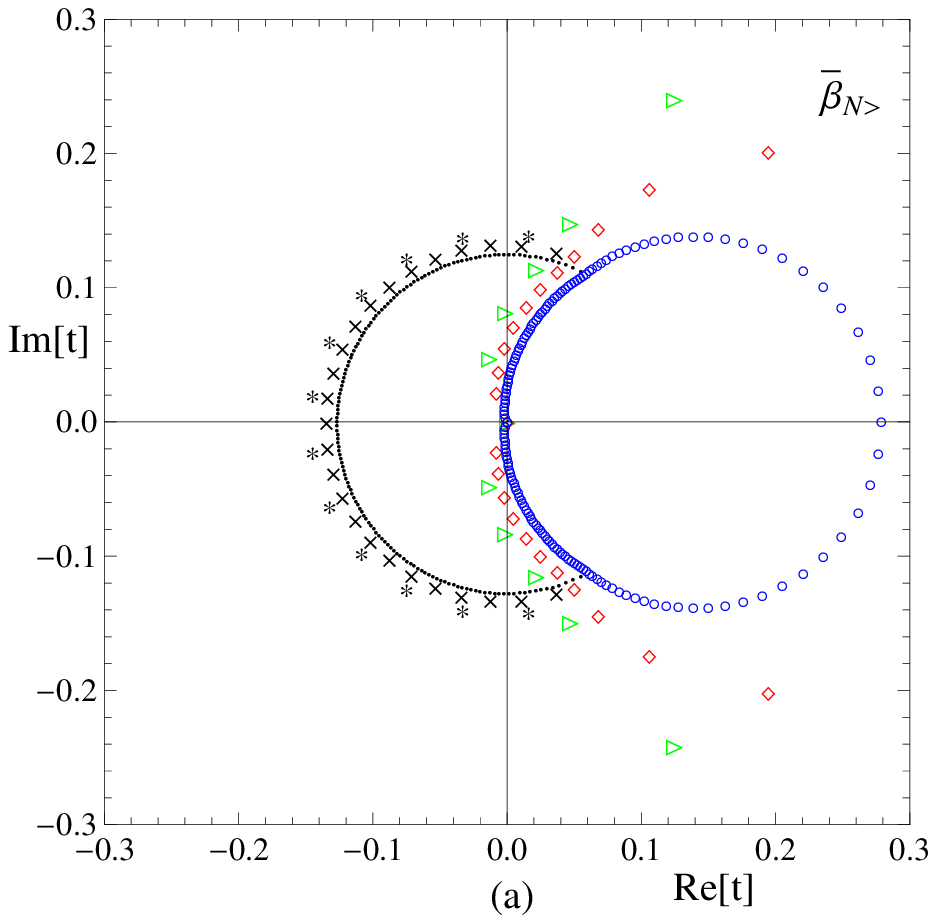}
\includegraphics[scale=0.85]{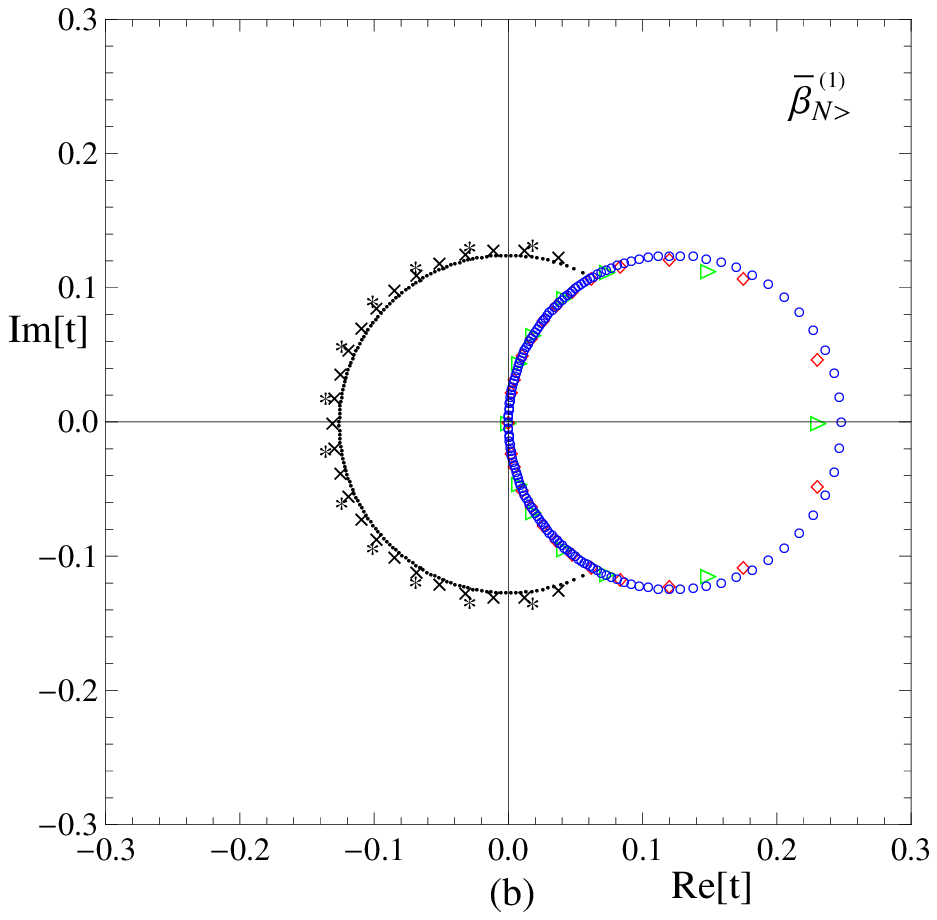}
\caption{(Color online) (a) Poles and zeros in diagonal Pad\'e approximants of $\bar\beta_{N>}$ in the complex $t$ plane.  Stars, crosses and dots in black for poles at $N=24$, $50$ and $300$ respectively.  Green triangles, red diamonds and small blue circles for zeros at $N=24$, $50$ and $300$ respectively.  (b) Same plots for $\bar\beta^{(1)}_{N>}$}
\end{figure}

From the argument so far, 
though it is hard to analytically proves the convergence in the large $N$ limit, we can say that there exists enough evidence of (\ref{delta_limit}), stated in this case as
\begin{equation}
\lim_{N\to \infty}\bar\beta_{N>}=\beta_{c}\quad t\in I=(0,1/4).
\end{equation}
Similarly, we conclude that 
\begin{equation}
\lim_{N\to \infty}\bar\beta_{N>}^{(\ell)}=0\quad t\in I\,\,(\ell=1,2,3,\dots).
\end{equation}
and hence for arbitrary $L$ and parameters $\rho_{\ell}$,
\begin{equation}
\lim_{N\to \infty}(\bar\beta_{N>}+\sum_{\ell=1}^{L}\rho_{\ell}\bar\beta_{N>}^{(\ell)})=\beta_{c}\quad t\in I.
\label{fundamental_theorem}
\end{equation}
An example of the above kind of combinations is $\bar\beta_{N>}+2\bar\beta^{(1)}_{N>}$ considered before corresponding $L=1$ and $\rho_{1}=2$.   The speed of convergence as $N$ grows  depends on the value of $\rho_{\ell}$.  From (\ref{crit_2d_a}), we find that the suitable set is $(\rho_{1}=1/2\nu, \rho_{2}=1/2\nu+1, \cdots)$ but the spectrum of the true exponents is the target itself in our study and not allowed to be used as an input.  In the next section we discuss how to use (\ref{fundamental_theorem}) for the estimation of critical temperature and exponents.

\subsection{Estimation of $\nu$ and $\beta_{c}$}
Figure 3(b) clearly exhibits that $\bar\beta_{N>}$ recovers the scaling behavior.  This provides us the possibility of estimating critical quantities from $\bar\beta_{N>}$ effective at small $t$.   In this subsection, we explicitly carry out estimations of critical quantities $\beta_{c}$ and $\nu$.  
In fact, $\beta_{c}$ estimation is able to carry through with the protocol presented in Ref. \cite{yam}.  Though we shall revise the protocol later, we first present the protocol and the result below.

We start with the scaling form written in $x$,
\begin{equation}
\beta_{<}=\beta_{c}-A_{1}x^{-p_{1}}-A_{2}x^{-p_{2}}-\cdots+R,
\label{ansatz}
\end{equation}
where $R$ denotes the analytic part (see (\ref{crit_2d})).  Consider the simplest ansatz to the first correction $\beta_{<}=\beta_{c}-A_{1}x^{-p_{1}}+R$.  From the argument on the $\delta$-expansion of regular series presented in the one-dimensional model, we neglect $D_{N}[R]$ and employe the simple ansatz,
\begin{equation}
\bar\beta_{N<}=\beta_{c}-\bar A_{1}t^{-p_{1}},\quad \bar A_{1}=A_{1}C_{N,-p_{1}}.
\end{equation}
The terms have exponents $0$, $-p_{1}$ and 
the above ansatz satisfies second order LDE $(0+d/d\log t)(p_{1}+d/d\log t)\bar\beta_{N<}=0$, and then provides after the integration over $\log t$
\begin{equation}
\Big[1+p_{1}^{-1}\frac{d}{d\log t}\Big]\bar\beta_{N<}=\beta_{c}.
\label{lde1}
\end{equation}
The inverse critical temperature appears as the integration constant.  
Then, we assume the basic result of $\delta$-expansion which states 
\begin{equation}
\lim_{N\to \infty}\Big[1+p_{1}^{-1}\frac{d}{d\log t}\Big]\bar\beta_{N>}=\beta_{c}\quad t\in I.
\label{base1}
\end{equation}
Note that $\bar\beta_{N>}$ effective at small $t$ is used in the place of $\bar\beta_{N<}$.  

Now, see Fig. 7 where $\bar\beta_{N>}$ and its derivatives are plotted at $N=25$.   We confirm from Fig. 3(a) and Fig. 7 that 
approximate scaling behaviors for $\bar\beta$ and $\bar\beta^{(1)}$ can be observed in the region $t\in I=(0.2, 0.23)$.   (We see from (\ref{ansatz}) that $\beta_{<}^{(1)}=p_{1}A_{1}x^{-p_{1}}+p_{2}A_{2}x^{-p_{2}}+\cdots+R^{(1)}$.  The decreasing trend of the first derivative is clearly exhibited in $\bar\beta_{25>}^{(1)}$)  
\begin{figure}
\centering
\includegraphics[scale=0.9]{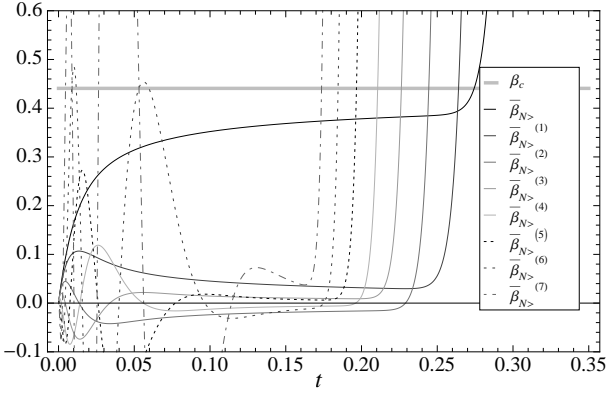}
\caption{$\bar\beta_{N>}(t)$ and its derivatives $\bar\beta_{N>}^{(\ell)}$ for $\ell=1,2,3,4,5,6,7$ at $N=25$.  The thick gray line represents $\beta_{c}=\log(1+\sqrt{2})/2$.   Though the scaling of $\bar\beta_{N>}^{(7)}$ has just begun to be observed, other derivatives exhibit scaling after the last bumps.  This allows us, at $25$th order, to use a three-parameter ansatz.  A four- or more multiparameter ansatz needs higher order terms above $25$th.}
\end{figure}

Then, we make use of the approximate case of (\ref{base1}) without the $N\to \infty$ limit;   that is, we identify the value of $[1+p_{1}^{-1}\frac{d}{d\log t}]\bar\beta_{N>}$ at a stationary point   $t=t^{*}$ in the plateau (see the plots in Fig. 4).  The result can be written as
\begin{equation}
\Big[1+p_{1}^{-1}\frac{d}{d\log t}\Big]\bar\beta_{N>}\Big |_{t=t^{*}}=\beta_{c}.
\label{recipe1}
\end{equation}
Then, the task remaining is to find most reliable or optimal 
 set of $t$ and $p_{1}$, which is expected to be close to $1/2\nu$.  For this purpose, the extension of the principle of minimum sensitivity (PMS) due to Stevenson \cite{steve} was used.  It suggests that the optimal set of $p_{1}$ and $t$ is determined by the following simultaneous equations
\begin{eqnarray}
\Big[1+p_{1}^{-1}\frac{d}{d\log t}\Big]\bar\beta_{N>}^{(1)} &=&0 \label{pms1}\\
\,\Big[1+p_{1}^{-1}\frac{d}{d\log t}\Big]\bar\beta_{N>}^{(2)} &\sim&0.
\label{pms2}
\end{eqnarray}
Here the symbol $\sim$ means the left-hand side must be vanishing or a local minimum in the  absolute value.  

Some comments would be in order:   In general, the solution of (\ref{pms1}) and (\ref{pms2}) is not unique at a given order, and among those solutions, the best one must be selected somehow.  Let us explain this issue by focusing on typical order cases, $N=9, 10,11$.  From (\ref{pms1}), we obtain $p_{1}$ as a function of a stationary point $t$ as $p_{1}^{-1}(t)=-\bar\beta_{N>}^{(2)}/\bar\beta_{N>}^{(1)}$.  For instance, at $N=9$, one has
\begin{equation}
\frac{1}{p_{1}(t)}=\frac{-9+288t-4872t^2+\cdots-9107446t^8}{9(1-64t+1624t^2-\cdots+9107446t^8)}.
\end{equation}
Substitution of $p_{1}^{-1}(t)$ into the left-hand-side of (\ref{pms2}) gives $[1+p_{1}^{-1}(t)d/d\log t]\bar\beta_{N>}^{(2)}(t)$ which is plotted in Fig. 8. 
\begin{figure}[h]
\centering
\includegraphics[scale=0.9]{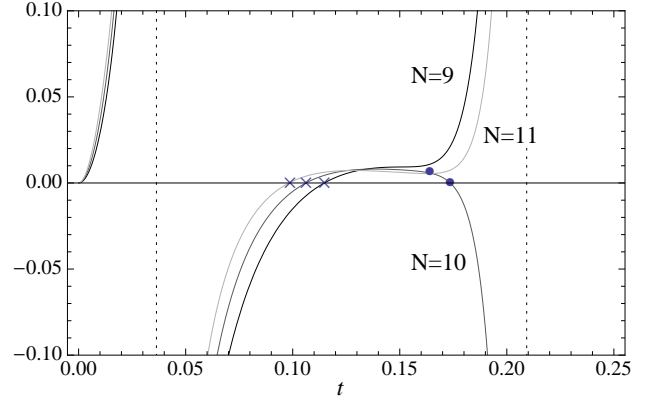}
\caption{$[1+p_{1}^{-1}(t)d/d\log t]\bar\beta_{N>}^{(2)}$ at $N=9,10,11$.  The dotted vertical lines indicate the singularities.}
\end{figure}
The optimal solution is determined in the following way:  At $N=9$ we have just one solution at the cross point and this gives optimal $t^{*}=0.11477$.   At $N=10$ there is one more solution depicted by the black blob.   The peak between the two zeros shows the turning point to the scaling and therefore the blob point is plausible to be selected as the optimal one, giving $t^{*}=0.173574$.  Similarly at $N=11$, the local minimum represented by blob point corresponds to the optimal one and $t^{*}=0.16404$.   Then we obtain the optimal estimation of $p_{1}$ by $p_{1}(t^{*})=p_{1}^{*}$ and the estimation of $\beta_{c}$ is given by (\ref{recipe1}).  We use the term "proper solution" for those depicted by the filled blobs at $N=10$ and $11$.  Just from these orders, the proper approximation sets in.  At $N=9$, the set $(t^{*},p_{1}^{*})$ is optimal but not proper.

We can also carry out multi-parameter or $K$th order LDE analysis from
\begin{equation}
\prod_{i=1}^{K}\Big[1+p_{i}^{-1}\frac{d}{d\log t}\Big]\bar\beta_{N>}=\beta_{c}.
\label{cond}
\end{equation}
Extending the PMS idea, we require that $\bar\beta_{N>}$ satisfies at some $t$ that
\begin{eqnarray}
\prod_{i=1}^{K}\Big[1+p_{i}^{-1}\frac{d}{d\log t}\Big]\bar\beta_{N>}^{(1)}&=& 0,\label{cond6}\\
\cdots &\cdots &\nonumber\\
\prod_{i=1}^{K}\Big[1+p_{i}^{-1}\frac{d}{d\log t}\Big]\bar\beta_{N>}^{(K)}&=& 0,\label{cond7}\nonumber\\
\prod_{i=1}^{K}\Big[1+p_{i}^{-1}\frac{d}{d\log t}\Big]\bar\beta_{N>}^{(K+1)}&\sim& 0.\label{cond8}
\end{eqnarray}
The first $K$ condition allows a solution $p_{1}$, $p_{2}$, $\cdots$, $p_{K}$ as functions of $t$ which specifies the estimation point.   Then, the last condition (\ref{cond8}) gives the optimal or proper solution for $t=t^{*}$, and the substitution gives the estimate $p_{i}^{*}=p_{i}(t^{*})$.  Estimation of $\beta_{c}$ can then be done with (\ref{cond}).
\begin{table*}
\caption{$K$-parameter estimation results of $\beta_{c}=0.44068679\cdots$ ($K=1,2,3,4,5$).  At $K=1$, a proper estimate appears from the $8$th for even orders and the $11$th for odd orders.  At $K=2$, a proper estimate appears from the $14$th for even orders and the $17$th for odd orders.  At $K=3$, a proper estimate appears from the $22$nd for even orders and the $25$th for odd orders.  The estimate before the appearance of a proper one is labeled by an asterisk.}
\begin{center}
\begin{tabular}{lccccccccc}
\hline\noalign{\smallskip}
   & $10$ & $15$ & $20$ & $25$ & $30$ & $35$ & $40$ & $45$ & $50$\\
\noalign{\smallskip}\hline\noalign{\smallskip}
$K=1$  & 0.43451427  & 0.43768742  &  0.43905467  &  0.43952324 & 0.43989324  & 0.44004633 & 0.44020306 & 0.44027202 & 0.44035468\\
$K=2$  & 0.44270440$^*$  & 0.43995506 &  0.44045465 &  0.44057056 & 0.44062720 & 0.44064773 & 0.44066261 & 0.44066888 & 0.44067449  \\
$K=3$  &    &  &  0.44064208$^*$ &  0.44065581 & 0.44067626 & 0.44068146 & 0.44068414 & 0.44068514 & 0.44068583  \\
$K=4$  &    &  &    &    & 0.44068347 & 0.44068540 & 0.44068627 & 0.44068653 & 0.44068666 \\
$K=5$  &    &  &    &    &  & 0.44068684$^*$ & 0.44068662 & 0.44068672 & 0.44068677 \\
\noalign{\smallskip}\hline
\end{tabular}
\end{center}
\end{table*}

In general, as many terms are included into the ansatz, the accuracy of estimate is increased, though the onset of proper estimation is delayed to higher orders.  See fthe numerical results summarized in Table I.  An approximate assessment at which order $\bar\beta_{N>}$ begins to show that scaling is found from the plot of the function and its derivatives.  For instance, at $N=25$, we notice that, though the seventh-order derivative which enters the job at a three-parameter ansatz shows rough scaling, all the derivatives can be said to be exhibiting the scaling behaviors.  It is also notable that from the plot one can guess the value of proper solution $t^{*}$.  On these grounds, one may expect that the three parameter ansatz is available at $N=25$.  More precisely, we consider that reliable estimation begins with the emergence of the proper stationary point.  And, at a given order of large $M$ expansion, the best approximation of $\beta_{c}$ is obtained in an ansatz with $K$ parameters where $K$ denotes the maximal number of parameters providing a proper solution $t^{*}$.   This is the summary of the protocol used in Ref. \cite{yam}.

The estimate of an exponent $p_{i}$ by $p_{i}^{*}$ accepts slightly less accuracy.  For instance at $N=25$, a three-parameter ansatz meets just proper case and yields $p_{1}^{*}=0.5007(\nu=0.9986)$, $p_{2}=1.6537$.   At $N=50$, a proper six-parameter ansatz yields $p_{1}^{*}=0.5000006(\nu=0.9999989)$, $p_{2}=1.5011444$, and $p_{3}=2.565879$.  The value of $p_{2}$ at $50$th order proves that the next order term is $M^{3/2}$ and not $M^{1}$, which is contained in $R$ (see (\ref{crit_2d})).  The value of $p_{3}$ at $N=50$ indicates the next-to-the-next order is $M^{5/2}$.   All of these results are in accord with (\ref{regular}), (\ref{beta_scale1}), and (\ref{beta_scale2}).  

In the previous protocol, critical exponents are estimated in the relation with $\beta_{c}$.  Here we try to estimate the leading exponent $p_{1}$ in the manner independent of $\beta_{c}$ estimation.   The point is simple.  It just suffices to differentiate $\beta(t)$ by $\log t$ to remove $\beta_{c}$ and then construct a quantity which converges to $p_{1}$ in the $M\to \infty$ limit.   We will demonstrate that the ratio of the derivatives $f_{\beta}=\beta^{(2)}/\beta^{(1)}$ is convenient for the aim and the estimation accuracy will be improved.  One remark is in order here.  In working $\beta(x)$, we assumed that the regular part $R$ is negligible and the numerical results on the exponents $p_{2}$ and $p_{3}$ at high orders support that it is the case.  However, in $f_{\beta}$, a regular part plays a role of effective corrections as we can see below:  Writing $\beta_{<}^{(\ell)}=-\sum_{n=1}A_{n}(-p_{n})^{\ell}x^{-p_{n}}+R^{(\ell)}$, we find 
\begin{eqnarray}
f_{\beta<}&=&\{-p_{1}-(A_{2}/A_{1})(p_{2}/p_{1})(p_{2}-p_{1})x^{p_{1}-p_{2}}+\cdots\}\nonumber\\
& &+\{x^{p_{1}}(R^{(1)}/A_{1}+R^{(2)}/(A_{1}p_{1}))+\cdots\}.
\label{ratio2d}
\end{eqnarray}
The first part implies the expansion result comes solely from non-analytic part of $\beta(x)$, while the last part involves $R$.  We find that $R$ dependent part has expansions in the fractional powers of $x$.  On the other hand, actually, the first part is a regular series, and the $\delta$-expansion would make it be negligible except for the first term $-p_{1}$.  Thus $R$ plays a major role in $f_{\beta}$.

Neglecting higher order corrections, $f_{\beta}$ satisfies
\begin{equation}
\prod_{i=1}^{K_{\nu}}\Big[q_{i}+t\frac{d}{dt}\Big]\bar f_{\beta N<}= 0,\label{dif1}
\end{equation}
where $q_{1}=0$, indicating the constant term $-p_{1}$ in (\ref{ratio2d}).  Another exponent $q_{i}$ $(i=2,3,4,\cdots)$ stands for the power of terms involved in the second part of (\ref{ratio2d}).   Here, $K_{\nu}$ stands for the number of terms included in $\bar f_{\beta N <}$.  Integration with respect to $t$ and the replacement $\bar f_{\beta N<}\to \bar f_{\beta N>}$ would then give
\begin{equation}
\lim_{N\to \infty}\prod_{i=2}^{K_{\nu}}\Big[1+q_{i}^{-1}t\frac{d}{dt}\Big]\bar f_{\beta N>}= -p_{1}.
\label{dif2}\\
\end{equation}
This is the equation we start with.  Conditions to estimate $-p_{1}$ are obtained by the differentiation, giving
\begin{eqnarray}
\prod_{i=2}^{K_{\nu}}\Big[1+q_{i}^{-1}t\frac{d}{dt}\Big]\bar f_{\beta N>}^{(1)}&=& 0,
\label{cond9}\\
\prod_{i=2}^{K_{\nu}}\Big[1+q_{i}^{-1}t\frac{d}{dt}\Big]\bar f_{\beta N>}^{(2)}&=& 0,\label{cond10}\\
\cdots &\cdots & \nonumber\\
\prod_{i=2}^{K_{\nu}}\Big[1+q_{i}^{-1}t\frac{d}{dt}\Big]\bar f_{\beta N>}^{(K)}&\sim& 0.\label{cond11}
\end{eqnarray}
The estimation details are same as those in the estimation of $\beta_{c}$.  The only thing worth mentioning is that the series of $\bar f_{\beta N>}$ becomes somewhat less-qualified compared to the series of $\bar\beta$ itself.  This depletion is reflected in the derivatives of $\bar f_{\beta N>}$, which is depicted in Fig. 9.  For the $1$-LDE case, a proper sequence appears at eighth order for even orders and $11$th for odd orders.  For the $2$-LDE sequence, a proper estimate begins from $16$th for even orders and $19$th for odd orders.  For the 3-LDE sequence, we have $26$th and $29$th orders, and for the $4$-LDE sequence, $36$th and $39$th orders.  The result of the estimate is summarized in Table II and depicted in Fig. 10.
\begin{figure}
\centering
\includegraphics[scale=0.9]{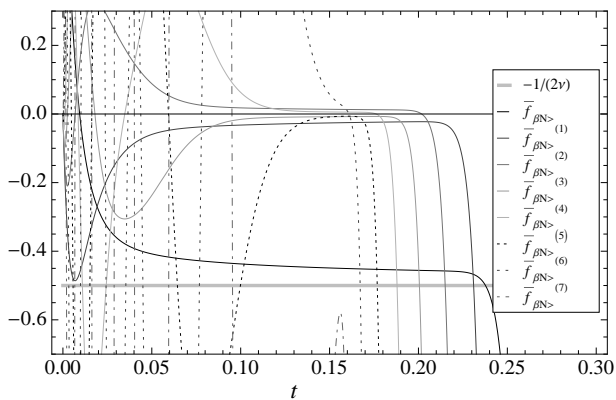}
\caption{Derivatives of $\bar f_{\beta N>}$ at $N=25$.  The seventh-order derivative shows behavior far from scaling and the sixth-order derivative is just about setting in the scaling.  The thick gray line indicates $-p_{1}=-1/(2\nu)=-0.5$.}
\end{figure}
\begin{figure}
\centering
\includegraphics[scale=0.9]{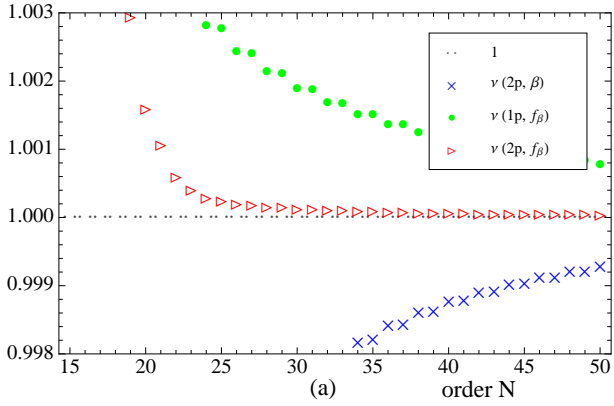}
\includegraphics[scale=0.9]{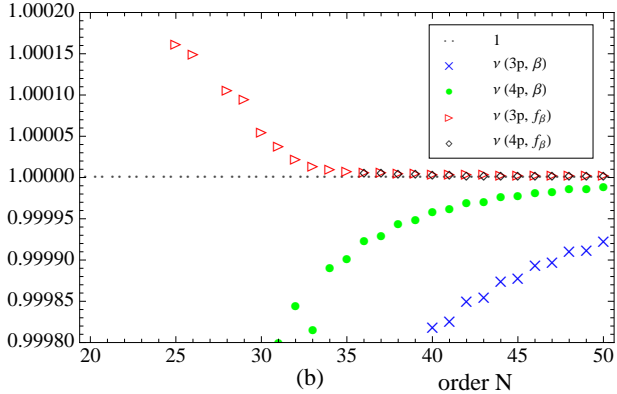}
\caption{(Color online) Estimated $\nu$ via $\beta$ and $f_{\beta}$.  "$\nu (Kp,\beta)$" means the result from the $K$-parameter ansatz of $\beta_{<}$.  Similarly,  "$\nu (Kp, f_{\beta})$" means the result from the $K$-parameter ansatz of $f_{\beta <}$.  Plot (a) shows the results via the one- and two-parameter ansatz and plot (b) via the three- and four-parameter ansatz.  The dotted lines indicate $\nu=1$.  Both of (a) and (b) show that the estimations via $\bar f_{\beta}$ are superior to those via $\bar \beta$.}
\end{figure}
\begin{table*}
\caption{Estimation results of $\nu$ at orders $N=15, 20, \cdots, 40$.  The results via $\bar\beta$ are listed in the column labeled by the values of $K$, which stands for the number of parameters involved in LDE for $\bar\beta$.   The results via $\bar f_{\beta}$ are shown in the column labeled by $K_{\nu}$, which stands for the number of parameters involved in LDE for $\bar f_{\beta}$.   In the later improved estimate, proper results appear from the 16th order in the $K_{\nu}=2$ case, and the 26th order in the $K_{\nu}=3$ case.  Results before the appearance of a proper sequence are indicated by an asterisk.}
\begin{center}
\begin{tabular}{lcccccc}
\hline\noalign{\smallskip}
& $15$ & $20$ & $25$ & $30$ & $35$ & $40$ \\
\noalign{\smallskip}\hline\noalign{\smallskip}
$K_{\nu}=1$  & 1.0072333  & 1.0039272  &  1.0027581  &  1.0018766 & 1.0014966  & 1.0011272 \\
$K=2$  & 0.9829895  & 0.9928665 &  0.9958010 &  0.9974954 & 0.9981953  & 0.9987589 \\
$K_{\nu}=2$  & 1.0051863$^*$  & 1.0015675  &  1.0002212  &  1.0001129 & 1.0000743  & 1.0000465 \\
$K=3$  &    &   & 0.9986069  & 0.9994266 & 0.9996737 & 0.9998178  \\
$K_{\nu}=3$  &    &   &  1.0001608$^*$ &  1.0000539 & 1.0000069 & 1.0000033  \\
$K=4$  &    &   &    & 0.9997888 & 0.9998998 & 0.9999569  \\
$K_{\nu}=4$  &    &   &    &    & 1.0000048$^{*}$ & 1.0000018  \\
\noalign{\smallskip}\hline
\end{tabular}
\end{center}
\end{table*}

In Table II and Fig. 10, the results of the $p_{1}$ estimate by the previous protocol in the 2-LDE ansatz are also plotted.  As a typical case, compare 2-LDE results with the improved 2-LDE (including $q_{2}$ and $q_{1}$) results.   We see from Table II that the improved 2-LDE results are superior to the original 2-and 3-LDE results when those are available at a given order.   The only drawback in the former is the delay of the emergence of the proper sequence.   The correction exponents can be also estimated:  At $N=50$, a proper five-parameter ansatz is available and yields $q_{2}=0.500002$, $q_{3}=1.502126$, and $q_{4}=2.594018$.  This result serves as evidence that $q_{2}=1/2$, $q_{3}=3/2$, and $q_{4}=5/2$.  Note that the integer power is not detected.
 
\begin{table*}
\caption{Improved $K$-parameter estimation results of $\beta_{c}=0.44068679\cdots$ ($K=2,3,4$).  $K=2$ results are from $K_{\nu}=2$ estimate of $\nu$ and $K=3,4$ results are from $K_{\nu}=3$ estimate of $\nu$.  Results before the appearance of proper sequence is indicated by the asterisk.}
\begin{center}
\begin{tabular}{ccccccc}
\hline\noalign{\smallskip}
& $15$ & $20$ & $25$ & $30$ & $35$ & $40$ \\
\noalign{\smallskip}\hline\noalign{\smallskip}
$K=2$  & 0.44104211  & 0.44079548   &  0.44071637   &  0.44070377  & 0.44069666   & 0.44069360 \\
$K=3$  &    &   &  0.44069402 &  0.44068944 & 0.44068762 & 0.44068726  \\
$K=4$  &    &   &  0.44069045$^*$  &  0.44068811 & 0.44068702 & 0.44068689  \\
$K=5$  &    &   &    &   & 0.44068692 & 0.44068685  \\
\noalign{\smallskip}\hline
\end{tabular}
\end{center}
\end{table*}
By using the $p_{1}$-estimate obtained in this manner, the improved estimate of $\beta_{c}$ becomes possible as we can see below:   The new protocol tells that, at a given order, one should use the ansatz of $\bar\beta_{N<}(t)$ with maximal number of parameters, say $K$.  Then we bias $\beta_{c}$ estimation via $\bar\beta_{N>}(t)$ with $K$-LDE by the substitution of $p_{1}^{*}$  obtained in the $K_{\nu}$-LDE estimation via $f_{\beta}$, where the order $K_{\nu}$ stands for the available maximal order of LDE.   For instance, at $25$th order a three-parameter ansatz is available (see Table I).  Then, we can use $K_{\nu}=2$ proper estimate of $p_{1}$ via $f_{\beta}$.  On the other hand, at $26$th order, we deal with the same three-parameter ansatz with $K_{\nu}=3$ proper estimate of $p_{1}$.  At each orders, we substitute $p_{1}$ obtained in respective orders into the set of conditions (\ref{cond6}) - (\ref{cond8}).  Since the $p_{1}$ exponent is thus fixed, one can discard one condition.  The last condition (\ref{cond8}) involves the highest derivatives among them and suitable to be discarded.  In this way, we redo the estimation of $\beta_{c}$, and the results are summarized in Table III and Fig. 11.

\begin{figure}
\centering
\includegraphics[scale=0.9]{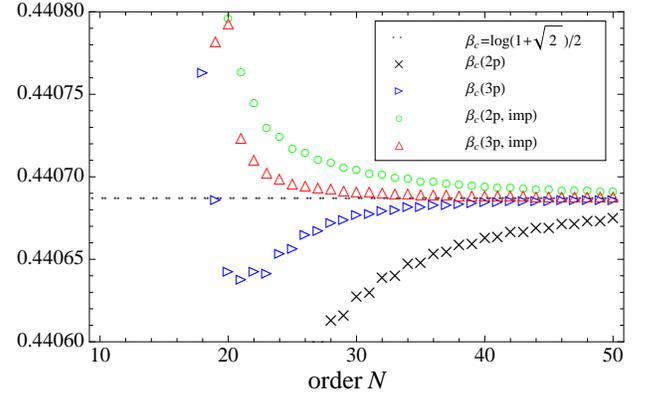}
\caption{(Color online) Estimation results of $\beta_{c}$ in the naive and improved protocols.   "$\beta (Kp)$" means the naive result from the $K$-parameter ansatz of $\beta_{<}$.  Similarly,  "$\beta (Kp, imp)$" means the improved result from $K$-parameter ansatz of $\beta_{<}$ under the bias of estimated $\nu$ via $f_{\beta}$.  The dotted line indicates $\beta_{c}$.  For a three-parameter ansatz, non-proper results are also plotted.}
\end{figure}

At $25$th order, for example, the naive estimation gives $\beta_{c}=0.4406558$ in $3$-LDE.  In the improved estimation, we have $\beta_{c}=0.4406940$ in $3$-LDE.  The improved result is more accurate than the naive one.  The accuracy improvement occurs to all orders we have examined to $50$th order.  We conclude that the new formula improves the accuracy of the $\beta_{c}$ estimate.

\subsection{Estimation of $\gamma/\nu$}
The exact and closed expression of the magnetic susceptibility defined by
\begin{equation}
\chi=\sum_{{\rm all\, sites}\, n}<s_{0}s_{n}>,
\end{equation}
is not obtained yet.  However, its mathematical structure has been vigorously investigated for some years.  References \cite{mc,orr,bou} provide good guidance.  It is now known that the behaviors of $\chi$ near the transition point read \cite{orr},
\begin{eqnarray}
\chi &=&C_{\chi} s^{-\gamma}(1+c_{1}s+c_{2}s^2+\cdots)\nonumber\\
& &+(\sqrt{1+s^2}+1)^{1/2}\Big\{(d_{0}+d_{1}s+\cdots)\nonumber\\
& &+\log s(e_{11}s+e_{12}s^3+\cdots)+(\log s)^2(e_{21}s^4+\cdots)\nonumber\\
& &+\cdots\Big\},
\label{chiscale}
\end{eqnarray}
where $s=(1/\sinh 2\beta-\sinh 2\beta)/2$ and
\begin{eqnarray}
C_{\chi}&=&1.41536651474451003415569965497895925122365318,\nonumber\\
c_{1}&=&\frac{1}{2},\quad c_{2}=\frac{5}{8},
\nonumber\\
d_{0}&=&-0.104133245093831026452160126860473433716236,\nonumber\\
d_{1}&=&-0.074368869753207080019958591697995003280476,\nonumber\\
e_{11}&=&0.032352268477309406090656526721221666637730,\nonumber\\
e_{21}&=&0.00939156987114587213179533187270757706.
\end{eqnarray}
Use of the expansion of $s$ in $M$, 
\begin{eqnarray}
s&=&\Big(\frac{M}{2}\Big)^{1/2}+\frac{1}{8}\Big(\frac{M}{2}\Big)^{3/2}-\frac{1}{128}\Big(\frac{M}{2}\Big)^{5/2}+\frac{181}{5760}\Big(\frac{M}{2}\Big)^{3}\nonumber\\
& &+\frac{7183}{322560}\Big(\frac{M}{2}\Big)^{7/2}+\frac{136459}{5160960}\Big(\frac{M}{2}\Big)^{4}+O(M^{9/2}),
\end{eqnarray}
then gives the behavior of $\chi$ described in $x$,
\begin{eqnarray}
\chi&=&C_{\chi}(2x)^{7/8}\Big\{1+c_{1}(2x)^{-1/2}+(c_{2}-\frac{7}{32})(2x)^{-1}\nonumber\\
& &+(c_{3}-\frac{3}{32})(2x)^{-3/2}+O(x^{-2})\Big\}\nonumber\\
& &+\Big\{d_{0}+(\frac{1}{2}d_{0}+d_{1})(2x)^{-1/2}+(\frac{1}{8}d_{0}+\frac{1}{2}d_{1}+d_{2})(2x)^{-1}\nonumber\\
& &+O(x^{-3/2})\Big\}\nonumber\\
& &+\Big\{-e_{11}(2x)^{-1/2}+O(x^{-3/2})\Big\}\log(2x)+\cdots.
\label{chi_scaling}
\end{eqnarray}

Compared with the critical behavior of $\beta(x)$, $\chi$ shows more complicated structure.  The spectrum of exponents comes in two families of $7/8-n/2\,(n=0,1,2,3,\cdots)$ and $0-n/2\,(n=0,1,2,3,\cdots)$.  In addition the later family allows logarithms which are considered as the remnant of the exponents with double, triple, etc degeneracies.   The similar but more complicated structure is observed in the ratio function $f_{\chi}=(\log\chi)^{(1)}=\chi^{(1)}/\chi$.  Though $f_{\chi}$ is more complicated than $\chi$, its leading term is just a constant $\gamma/2\nu$ and convenient for its estimation.  We therefore work with $f_{\chi}$.   First of all, 
we do not make use of the detailed result (\ref{chi_scaling}), since our purpose here is to assess the power of our approach to estimate the $\chi$-exponent $\gamma$ under simple assumptions.  We start in a robust manner with the supposed power-like behavior with analytic back ground,
\begin{equation}
\chi=\hat C_{\chi}\tau^{-\gamma}(1+\hat c_{1}\tau+\hat c_{2}\tau^2+\cdots)+R_{\chi},
\label{chi_scale}
\end{equation}
where $R_{\chi}=d_{0}+\hat d_{1}\tau+\cdots$. 
  Then, we have
\begin{equation}
f_{\chi}=\frac{\gamma}{2\nu}-\hat c_{1}p_{1}x^{-p_{1}}+\sum_{n=2}const\times x^{-r_{n}},
\label{fchi}
\end{equation}
where $p_{1}=1/2\nu$ and $r_{2}=\gamma/(2\nu)$.  The LDE to be satisfied by $\bar f_{\chi <}$ to the first $K$ expansion reads
\begin{equation}
\Big[0+\frac{d}{d\log t}\Big]\Big[p_{1}+\frac{d}{d\log t}\Big]\prod_{n=2}^{K}\Big[r_{n}+\frac{d}{d\log t}\Big]\bar f_{\chi N<}=0.
\end{equation}
Integration over $t$ and the replacement $\bar f_{\chi N<}\to \bar f_{\chi N>}$ then gives
\begin{equation}
\Big[1+p_{1}^{-1}\frac{d}{d\log t}\Big]\prod_{n=2}^{K}\Big[1+r_{n}^{-1}\frac{d}{d\log t}\Big]\bar f_{\chi N >}\sim \frac{\gamma}{2\nu}.
\label{lde_fchi}
\end{equation}
The behaviors of $\bar f_{\chi N>}^{(\ell)}$ $(\ell=0,1,2,\cdots)$ at $N=25$ are depicted in Fig. 12.  From the plots, we can understand roughly at which region of $t$, the estimation should be carried out in a given order.  At $N=25$ for example, all of $\bar f_{\chi N>}^{(\ell)}$ from $\ell=0\sim 5$ show approximate scaling around $t\sim 0.15$ and $t^{*}$ would be close to that.   It would be interesting to see whether the $\delta$-expansion plays an expected role in this case, i. e., whether $\bar f_{\chi N>}$ recovers the true complicated critical behavior.  We have numerically studied both functions $\bar f_{\chi N>}$ and $\bar f_{\chi N<}$ at several $N$ and confirmed it is the case.  For instance, Fig. 13 shows the plots of those at $N=25$.  The two functions are almost degenerate over a region of $t$ ranging from $\sim 0.15$ to $\sim 0.21$.  The agreement between them is confirmed also for the first and second derivatives at $N=25$.
\begin{figure}[t]
\centering
\includegraphics[scale=0.9]{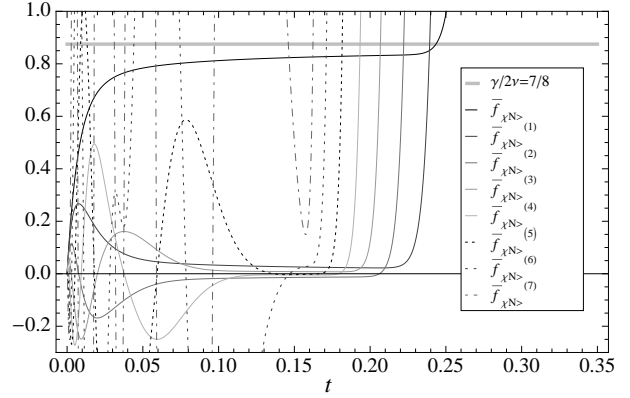}
\caption{Plots of $\bar f_{\chi N>}^{(\ell)}$ ($\ell=0,1,2,\cdots$) at $N=25$.  The thick gray line indicates $\gamma/(2\nu)=0.875$.}
\end{figure}
\begin{figure}[t]
\centering
\includegraphics[scale=0.9]{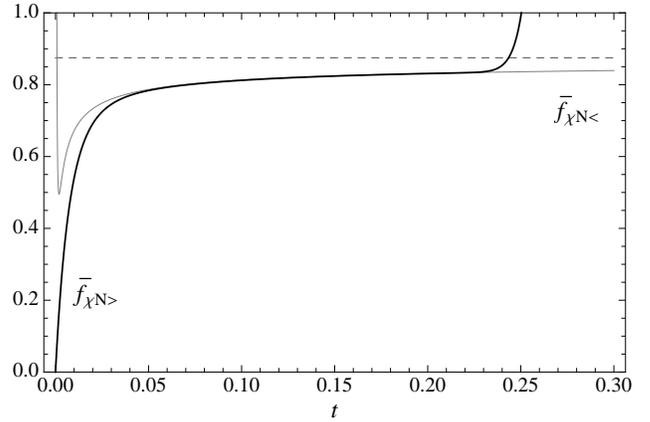}
\caption{Plots of $\bar f_{\chi N>}$ and $\bar f_{\chi N<}$ at $N=25$.  The dashed line indicates $\gamma/(2\nu)=0.875$.}
\end{figure}

In Ref. \cite{yam3}, an unbiased estimation of $\gamma$ via large mass expansion was done.  However, we must say that the result is not as good in the accuracy.    Here we note that the estimation of $\beta_{c}$ was improved under the bias of $\nu$ estimated order by order via $f_{\beta}$.   This would apply to the $\gamma$ estimation since the leading correction comes in $x^{-p_{1}}$.   We therefore employ the same procedure:  We substitute 
 $p_{1}=p_{1}^{*}$ estimated in the previous subsection into the left-hand-side of (\ref{lde_fchi}) and extended PMS conditions that follow from (4.28) to (4.30) under the replacement $\bar f_{\beta N>}\to \bar f_{\chi N>}$ in them. The result of $\gamma/2\nu$ estimation to $50$th order is shown in Table IV and Fig. 14.   It is apparent that the use of $p_{1}^{*}$ has made the accuracy better.  Here we notice that the used value of $p_{1}^{*}$  is the one obtained in the $1$-LDE for $f_{\beta}$.  The $k$-LDE with larger $k$ by which $p_{1}^{*}$ is supplied does not bring about any net improvement on $\gamma/2\nu$. 

\begin{figure}
\centering
\includegraphics[scale=0.9]{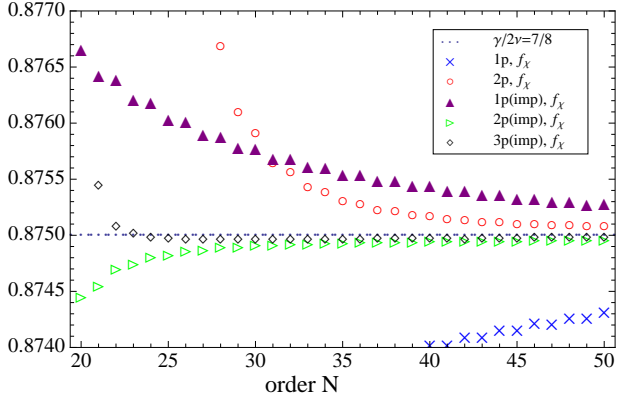}
\caption{(Color online) Estimation results of $\gamma/(2\nu)=7/8$ in the naive and improved protocols.   "$Kp,f_{\chi}$" means the unbiased result from the $K$-parameter ansatz of $f_{\chi <}$.  Similarly,  "$Kp(imp), f_{\chi}$" means the improved (biased) result from the $K$-parameter ansatz of $f_{\chi <}$..}
\end{figure}

\begin{table*}
\caption{Improved $K$-parameter estimation results of $\gamma/2\nu=7/8$ biased by $p_{1}^{*}$ estimated via $f_{\beta}$ order by order.}
\begin{center}
\begin{tabular}{cccccccc}
\hline\noalign{\smallskip}
& $20$ & $25$ & $30$ & $35$ & $40$ & $45$ & $50$ \\
\noalign{\smallskip}\hline\noalign{\smallskip}
$K=1$  & 0.87663881  & 0.87601885  &  0.87575751   &  0.87552991  & 0.87542884   & 0.87531581 & 0.87526554 \\
$K=2$  & 0.87444212  & 0.87481600  &  0.87490090 &  0.87492284 & 0.87493726 & 0.87494082 &  0.87494692 \\
$K=3$  &    & 0.87496338  &  0.87495516  & 0.87495917 & 0.87496394 & 0.87496794 &  0.87497436 \\
\noalign{\smallskip}\hline
\end{tabular}
\end{center}
\end{table*}
Our biased protocol may be said to be working basically well up to the moderate orders.  However, a complicated phenomenon occurs at higher orders.  For instance, a "second" proper solution appears about $N=68$ and $157$, respectively, in three- and four-parameter ansatze.  The region where the second proper solution lives is separated from the region of first proper solution by the pole of the relevant functions.   After the appearance of the second proper solution, the latter region gradually shrinks and the new region develops.  Thus, we come to be aware that the new solution must be kept as the true proper solution, and the old one should be called a pseudo proper solution.  As for one- and two- parameter ansatze, the second proper solution if it exists does not appear up to $N=500$, which is the maximum order we have studied.  However, a seemingly irregular phenomenon still occurs at higher orders.  For instance, using the exact value $p_{1}=1/2\nu=1/2$, we found that $[1+2\frac{d}{d\log t}]\bar f_{\chi N >}^{(1)}$ changes behavior at $N=350$.  See the change shown in Fig. 15.
\begin{figure}
\includegraphics[scale=0.9]{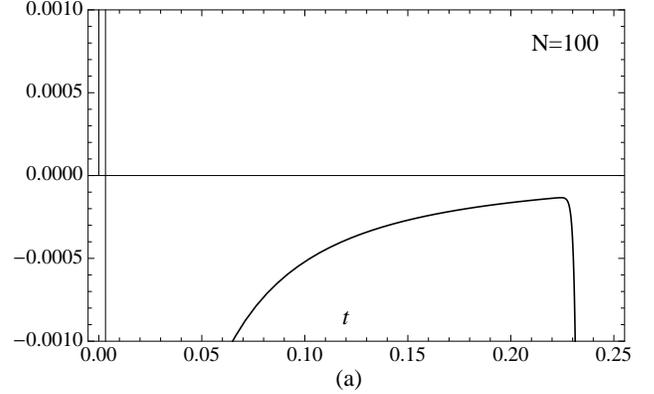}
\includegraphics[scale=0.9]{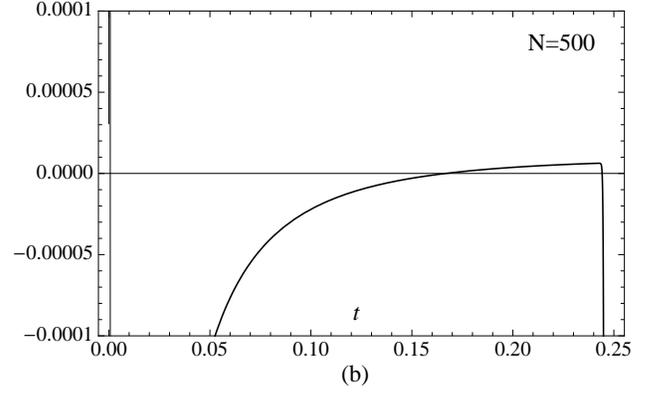}
\caption{$[1+2\frac{d}{d\log t}]\bar f_{\chi N >}^{(1)}$ at $N=100$ (a) and $500$ (b).}
\end{figure}
At $N=100$, $[1+2\frac{d}{d\log t}]\bar f_{\chi N >}^{(1)}$ approaches zero from below.  In contrast, at $N=500$, the same function goes across the zero-line at $t\sim 0.1667$.  As plotted in Fig. 16, from the exact critical behaviors to order $t^{-15/8}$ (the exponents of terms included are $0$, $-1/2$, $-7/8$, $-1$, $-11/8$, $-3/2$, $-7/4$, $-15/8$), we can observe similar behavior as in the case $N=500$ of $[1+2\frac{d}{d\log t}]\bar f_{\chi N >}^{(1)}$.   Though the matching of the plotted two functions is not good yet, the gray curve representing $[1+2\frac{d}{d\log t}]\bar f_{\chi N <}^{(1)}$ at $N=500$ passes across the zero line at $t\sim 0.08697$.  The value of the across point slightly increases as the number of terms included grows.  For instance, when the terms to $t^{21/8}$ are included, we have $t\sim 0.09631$ for the crossing point.  On the other hand, the evaluation of the crossing point in $[1+2\frac{d}{d\log t}]\bar f_{\chi N >}^{(1)}$ proves that $t=0.2082, 0.1852, 0.1667$ at $N=400, 450, 500$, respectively.    This would signal the reliability of the behavior of $[1+2\frac{d}{d\log t}]\bar f_{\chi N >}^{(1)}$ at quite large orders.  We emphasize that the discrepancy between the two functions is in fact small.  Indeed, the peak values are of the order $10^{-5}$, and the discrepancy is also of the same order.  The speed of approaching zero of $[1+2\frac{d}{d\log t}]\bar f_{\chi N >}^{(1)}$ is extremely slow.  This slow convergence to zero is a signal that this anomalous behavior originates from the logarithmic terms.
\begin{figure}
\includegraphics[scale=0.9]{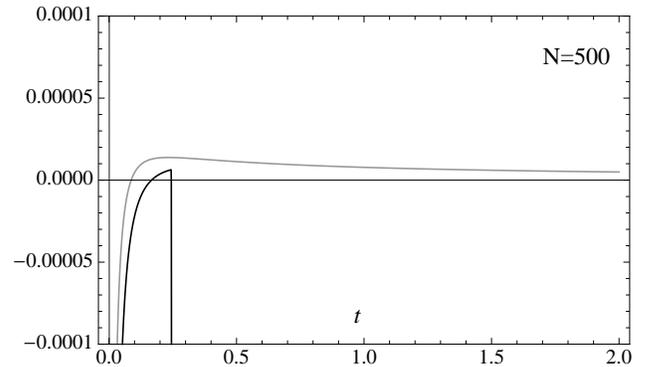}
\caption{$[1+2\frac{d}{d\log t}]\bar f_{\chi N >}^{(1)}$ at $N=500$ (black) and $[1+2\frac{d}{d\log t}]\bar f_{\chi N <}^{(1)}$ to $t^{15/8}$ (gray), in which $t^{11/8}\log t$ is included.}
\end{figure}

\begin{figure}
\includegraphics[scale=0.9]{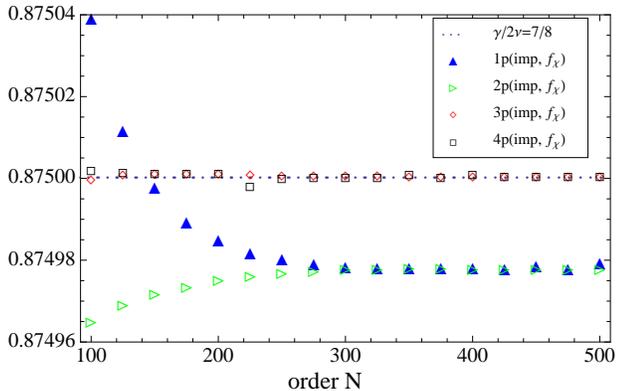}
\caption{(Color online) Estimation results of  for $\gamma/(2\nu)=7/8$ in the improved protocols at large orders.   "$Kp(imp), f_{\chi}$" means the result from the $K$-parameter ansatz of $f_{\chi <}$.}
\end{figure}
The sequences of one- and two- parameter ansatze seem to imply slightly smaller value of convergence limit (see Fig. 17).   However, we have checked that over $N\sim 400$, the estimate shows a slow increasing trend with small oscillation.  The reason of the appearance of an increasing trend even in the peak of the logarithmic effect is not seen in $[1+2\frac{d}{d\log t}]\bar f_{\chi N >}^{(1)}$ is that the coefficients of the logarithmic terms decrease as $N$ increases (see (\ref{scale_c1})).  

The same phenomenon does not occur in $\nu$ and $\beta_{c}$ estimations up to $N=500$.   This would be because $\bar\beta(t)$ and $\bar f_{\beta}(t)$ near $\beta_{c}$ are of the power-like structure.

\section{Concluding Remarks}
In the one-dimensional Ising model, the basic result (\ref{delta_limit}) is analytically proved.  In the square model, numerical work to large orders confirmed that (\ref{delta_limit}) is valid.  Based upon (\ref{delta_limit}), we performed the estimation of $\nu$, $\gamma$, and $\beta_{c}$  and demonstrated that the new protocol works better than the previous one \cite{yam}.  Also important is the confirmation of the convergence region of the $\delta$-expanded large $M$ expansion.  The existence is proved in one-dimension and verified numerically in the square model.  We emphasize in both cases that the convergence region is of the type $(0,t_{0})$, and the origin $t=0$ is excluded.

The presented approach has a characteristic property that one can estimate critical exponents directly from the high-temperature expansion, with no bias from the critical temperature.  Under the new protocol, we understand that $\nu$ can be estimated first and then, $\gamma$ and $\beta_{c}$ in the bias of estimated $\nu$.  It is logically natural that universal quantities are directly estimated by the large $M$ expansion.  The exponent $\nu$ has a special role in our approach because any typical thermodynamic quantity scales as $\sim \tau^{\gamma}(1+const\times \tau+\cdots)$ and the first correction has the extra power $1/2\nu$ due to the relation $\tau\sim M^{1/2\nu}$ near the critical point.  

In the traditional method such as DLog Pad\'e \cite{bak,gut2} and differential approximants methods \cite{gut,gut2} , $\beta_{c}$ is involved in the estimation equations.  Many estimation data are supplied in a given order and the statistical processing of those data is employed and the average becomes accurate compared to the raw data themselves.   In our method, in contrast, we have essentially a unique estimate.  Thus, there is no room for the statistical processing of the data.  The accuracy is not superior to those in the traditional approaches, but the logic is alternative and would be interesting since a direct path to the exponent is opened.


\begin{thebibliography}{99}
\bibitem{comment}  At small $\xi$, the two variable has connection $M\sim e^{1/\xi}$ and $M$ and $\xi^{-2}$ has no similarity.  Actualy, at small $\xi$ we have expansion of $\beta$ such as $\beta=e^{-1/\xi}+e^{-3/\xi}/3+x^{-5/\xi}/5+\cdots$.  The $\delta$-expansion is not applicable to this kind of series.  The squared mass $M$ cannot be replaced by $\xi^{-2}$.
\end{thebibliography}
\end{document}